\documentclass[preprint]{aastex631}
\usepackage{multirow}
\usepackage{enumerate}

\begin{document}

\title{The influence of thermonuclear bursts on polar caps of accreting X-ray millisecond pulsar MAXI J1816-195}

\author[0000-0001-9599-7285]{Long Ji\textsuperscript{*}}
\email{jilong@mail.sysu.edu.cn}
\affiliation{School of Physics and Astronomy, Sun Yat-sen University, Zhuhai, 519082, People's Republic of China}

\author[0000-0002-3776-4536]{Mingyu Ge\textsuperscript{*}}
\email{gemy@ihep.ac.cn}
\affiliation{Key Laboratory of Particle Astrophysics, Institute of High Energy Physics, Chinese Academy of Sciences, 19B Yuquan Road, Beijing 100049, China}

\author{Yupeng Chen\textsuperscript{*}}
\email{chenyp@ihep.ac.cn}
\affiliation{Key Laboratory of Particle Astrophysics, Institute of High Energy Physics, Chinese Academy of Sciences, 19B Yuquan Road, Beijing 100049, China}

\author[0000-0003-2310-8105]{Zhaosheng Li}
\affiliation{Key Laboratory of Stars and Interstellar Medium, Xiangtan University, Xiangtan 411105, Hunan, China}

\author{Peng-Ju Wang} 
\affiliation{Key Laboratory of Particle Astrophysics, Institute of High Energy Physics, Chinese Academy of Sciences, 19B Yuquan Road, Beijing 100049, People's Republic of China}

\author{Shu Zhang}
\affiliation{Key Laboratory of Particle Astrophysics, Institute of High Energy Physics, Chinese Academy of Sciences, 19B Yuquan Road, Beijing 100049, People's Republic of China}

\author{Shuang-Nan Zhang} 
\affiliation{Key Laboratory of Particle Astrophysics, Institute of High Energy Physics, Chinese Academy of Sciences, 19B Yuquan Road, Beijing 100049, People's Republic of China}

%% Note that the \and command from previous versions of AASTeX is now
%% depreciated in this version as it is no longer necessary. AASTeX 
%% automatically takes care of all commas and "and"s between authors names.

%% AASTeX 6.31 has the new \collaboration and \nocollaboration commands to
%% provide the collaboration status of a group of authors. These commands 
%% can be used either before or after the list of corresponding authors. The
%% argument for \collaboration is the collaboration identifier. Authors are
%% encouraged to surround collaboration identifiers with ()s. The 
%% \nocollaboration command takes no argument and exists to indicate that
%% the nearby authors are not part of surrounding collaborations.

%% Mark off the abstract in the ``abstract'' environment. 
\begin{abstract}
We report accretion-powered pulsations for the first time during thermonuclear bursts in hard X-rays, which were observed with \textit{Insight-HXMT} in 2022 during the outburst of {the} accreting X-ray millisecond pulsar MAXI J1816-195.
By stacking 73 bursts, we detected pulse profiles in 8-30\,keV and 30-100\,keV during bursts, which are identical to those obtained from the persistent (non-burst) emission.
On average, no significant phase lag was observed between burst and persistent pulse profiles.
In addition, we suggest that the interaction with burst photons can be used as a direct diagnostic to distinguish contributions from the hot plasma near polar caps and the corona around the accretion disk, which are highly degenerate in their spectral shapes.

\end{abstract}

%% Keywords should appear after the \end{abstract} command. 
%% The AAS Journals now uses Unified Astronomy Thesaurus concepts:
%% https://astrothesaurus.org
%% You will be asked to selected these concepts during the submission process
%% but this old "keyword" functionality is maintained in case authors want
%% to include these concepts in their preprints.

%% From the front matter, we move on to the body of the paper.
%% Sections are demarcated by \section and \subsection, respectively.
%% Observe the use of the LaTeX \label
%% command after the \subsection to give a symbolic KEY to the
%% subsection for cross-referencing in a \ref command.
%% You can use LaTeX's \ref and \label commands to keep track of
%% cross-references to sections, equations, tables, and figures.
%% That way, if you change the order of any elements, LaTeX will
%% automatically renumber them.
%%
%% We recommend that authors also use the natbib \citep
%% and \citet commands to identify citations.  The citations are
%% tied to the reference list via symbolic KEYs. The KEY corresponds
%% to the KEY in the \bibitem in the reference list below. 

\section{Introduction} \label{sec:intro}
Accreting millisecond X-ray pulsars (AMXPs) are weakly magnetized and fast-spinning
 neutron stars normally located in low mass X-ray binaries (LMXBs).
The accreted matter couples to magnetic lines around the magnetosphere, and eventually falls onto polar caps of the neutron star, resulting in X-ray pulsations.
They are spun up by the transferring of angular momentum from companion stars, i.e., the so-called  ``recycling scenario" \citep{Bhattacharya1991}.
Until now, dozens of AMXPs have been discovered and most of them are transient, showing faint outbursts on time scales of days to months \citep[for a review, see][]{Patruno2021}.
Their spectra can be described as a combination of one or two thermal components and a Comptonization component, similar to those observed in hard states of non-pulsing LMXBs which usually show spectral state transitions \citep[e.g.,][]{Gierlinski2005, Bozzo2010, Papitto2013, LiZS2023}.
The Comptonization component might originate from some hot plasma above the accretion disk (i.e., the ``corona") or/and the shock on polar caps of neutron stars.

Some AMXPs show thermonuclear bursts\footnote{They are also called ``Type-I bursts" in literature.} 
during their outbursts, which are triggered by unstable thermonuclear burning of accreted fuel on the neutron star surface \citep{Galloway2008, Galloway2021}.
They are characterized by a fast rise time followed by an exponential decay with a time scale of tens to hundreds of seconds.
Since their physics is relatively known, they are used as direct probes to study the accretion process by investigating the interactions between burst photons and materials surrounding the neutron star \citep{Degenaar2018}.
In observations, several processes have been studied through the changes of the persistent (non-burst) emission during bursts, including a hard X-ray shortage, an enhancement or a dip of the persistent emission and an additional reflection component \citep[e.g.,][]{Maccarone2003,Chen2012,Ji2014, Worpel2013, Worpel2015, Fragile2020, Speicher2023, Bult2021, Ballantyne2004, Keek2017, Keek2018,Zhao22,Lu23}.
We note that previous studies were mainly based on bursts in normal LMXBs because of their large number, while the situation in AMXPs might be different.
In MAXI J1816-195, \citet{Chen2022} reported 73 bursts detected with \textit{Insight-HXMT} satellite during the peak and decay phase of its 2022 outburst, making one of the largest burst sample of AMXPs.
They found only a slight deviation of burst spectra from the blackbody model, i.e., no significant soft X-ray excess as reported in most bright bursts of other sources,
and a hard X-ray (i.e., 30-100\,keV) deficit of $\sim$30\% during bursts, which is lower than other bursters in LMXBs.
This may be simply due to the fact that in AMXPs the accretion disk is truncated by the magnetosphere and relatively far away from the neutron star, so that the Poynting-Robertson drag that leads to the enhanced accretion and the cooling effect of the corona by burst photons are {less effective} \citep{Chen2022}.
{Another possibility is that in AMXPs a significant amount of radiation originates from polar caps, which might be less affected by bursts.}
{Up to now, the interactions between the burst emission and the accretion flow infalling along magnetic lines are still poorly understood, although some efforts have been made in different aspects \citep{Lovelace2007, Galloway2007, Altamirano2008,Cavecchi2022}.}
In this paper, we performed a timing analysis by stacking bursts in MAXI J1816-195 to investigate if the bursts will influence on polar caps.

MAXI J1816-195 is a millisecond pulsar with a spin frequency of 528.6\,Hz \citep{Bult2022} newly discovered during the 2022 outburst \citep{Negoro2022}.
Its type-I bursts have been extensively studied by using {\it NICER}, {\it NuSTAR} and \textit{Insight-HXMT} observations \citep{Mandal2023,Bult2022,Chen2022}.
Coherent timing analysis has been done using \textit{Insight-HXMT} and {\it NICER} data, and the exact timing solution is available \citep{LiZS2023}.
In addition, a transient 2.5\,Hz modulation was reported by \citet{LiPP2023}, which might be produced in an unstable corona.

\begin{figure}
\centering
\includegraphics[width=0.6\textwidth]{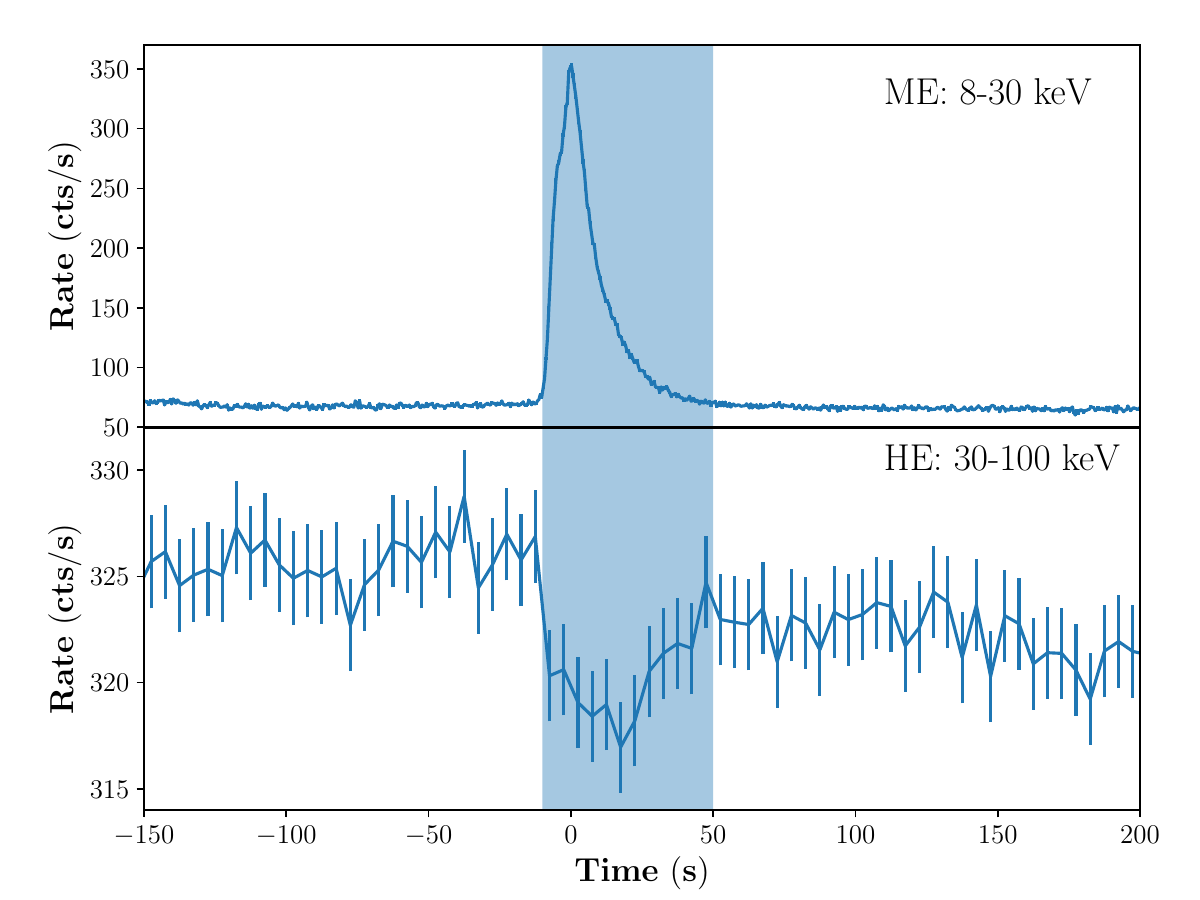}
\caption{Averaged lightcurves of thermonuclear bursts in MAXI J1816-195 observed with \textit{Insight-HXMT}/ME (upper) and HE (bottom) detectors.
The shaded region represents the burst interval.
}
\label{ME_HE_lc}
\end{figure}
\section{Data analysis and results}
We aim to compare burst and non-burst pulse profiles, since periodic signals can directly reflect the accretion near polar caps.
We note that this comparison can be performed more effectively in the hard X-ray ($\gtrsim$ 10\,keV) band, because thermonuclear bursts normally have a characteristic temperature of several keV, dominating soft X-rays during bursts and hampering the study related to the persistent emission.
The first Chinese X-ray astronomy satellite \textit{Insight-HXMT} \citep{Zhang2020} has an unprecedented effective area (5100\,$\rm cm^2$) in 20-250\,keV, which allows us to conduct this timing analysis in the hard X-ray band.
In this paper, we adopted the burst sample as reported by \citet{Chen2022}, which included 73 bursts observed with \textit{Insight-HXMT}/Medium Energy (ME) X-ray Telescope, among which 70 bursts were also found by the High Energy (HE) X-ray Telescope.
We analyzed the data using the software {\sc HXMTDAS v2.05}\footnote{\url{http://hxmtweb.ihep.ac.cn/software.jhtml}}, together with the calibration database {\sc CALDB v2.06}.
Following \citet{Chen2022}, we did not screen the raw data according to the officially recommended criteria, and instead only excluded the data during the South Atlantic Anomaly (SAA) passage.
We note that this has little influence on our results since the background does not present periodic modulation and has no contribution on the pulsed flux that we studied.
{
On the contrary, the non-pulsed flux of HE is dominated by the instrument background\footnote{As suggested by \citet{Chen2022}, on average the persistent emission from the source is about 40\,cts/s in the energy range of 30-100\,keV, while the background reaches up to $\sim$ 297\,cts/s.}.
}

For each observation containing bursts, we first extracted ME lightcurves with a bin size of 1\,s in the energy range of 8-30\,keV, and defined the burst peak time as the reference ($t=0$) in the following analysis.
We also defined the burst and non-burst time intervals as ``$-10<t<50$" and ``$-200<t<-50$ and $150<t<300$", respectively.
{During the burst interval, the prominent hard X-ray deficit has been reported by \citet{Chen2022}.
We have verified that a different definition of the burst interval has little influence on the results.}
Based on the reference time, we shifted and stacked all lightcurves of available bursts observed with ME and HE (Figure~\ref{ME_HE_lc}).
Clearly, during bursts there is an evident deficit shown in the averaged 30-100\,keV lightcurve.
We note that in 30-100\,keV the count rate is dominated by the background, and this deficit approximately corresponds to 30\% of the persistent emission before bursts \citep{Chen2022}.
In addition to the deficit, there is a downward trend after bursts.
It is most likely to be artificial due to an average change of the instrumental background  (see Appendix).

\begin{figure}
\centering
\includegraphics[width=0.6\textwidth]{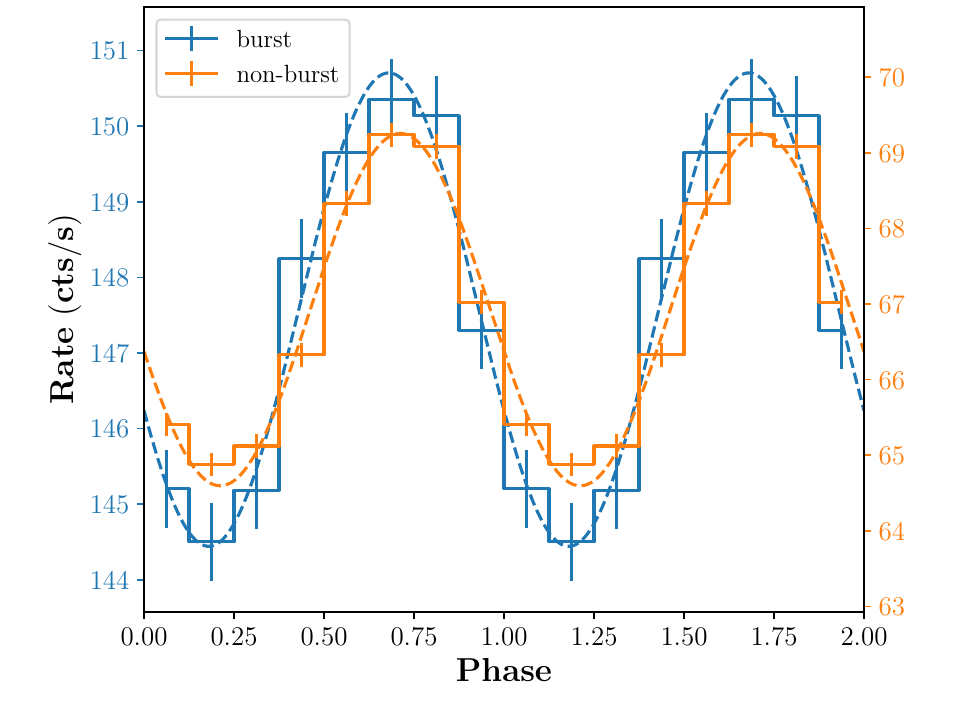 }
\includegraphics[width=0.6\textwidth]{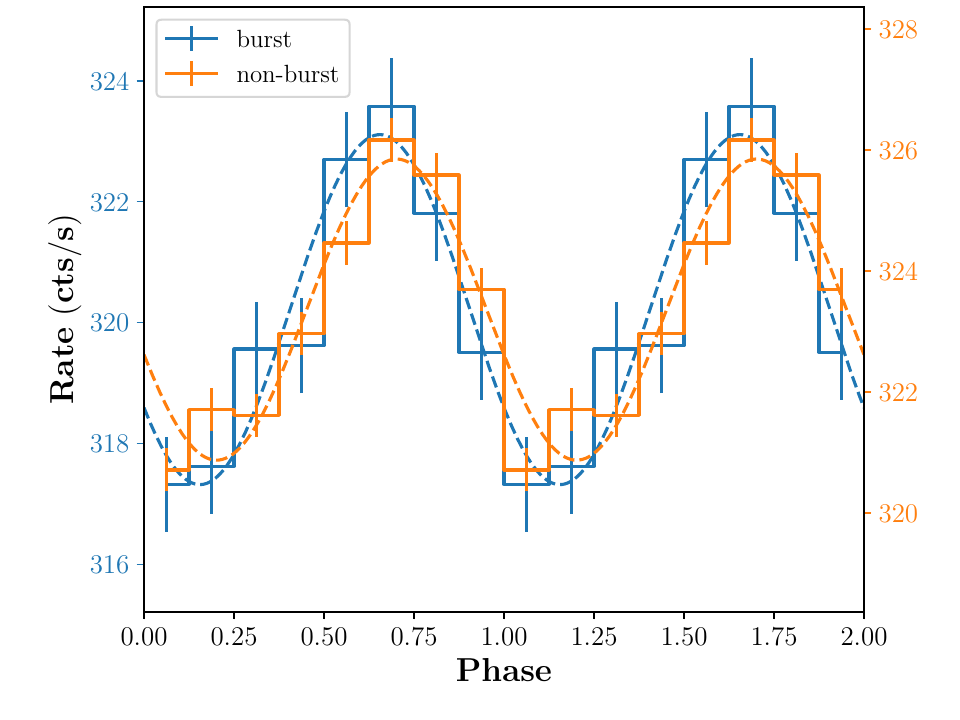 }
\caption{Averaged pulse profiles during thermonuclear bursts in 8-30\,keV (upper) and 30-100\,keV (lower), where dashed lines are best-fitting models assuming a sinusoidal shape.
The burst and non-burst pulse profiles correspond to left and right vertical axes, respectively.
We set the same scale for both axes, but with different  median values.
}
\label{profile_compare}
\end{figure}

\begin{figure}
\centering
\includegraphics[width=0.6\textwidth]{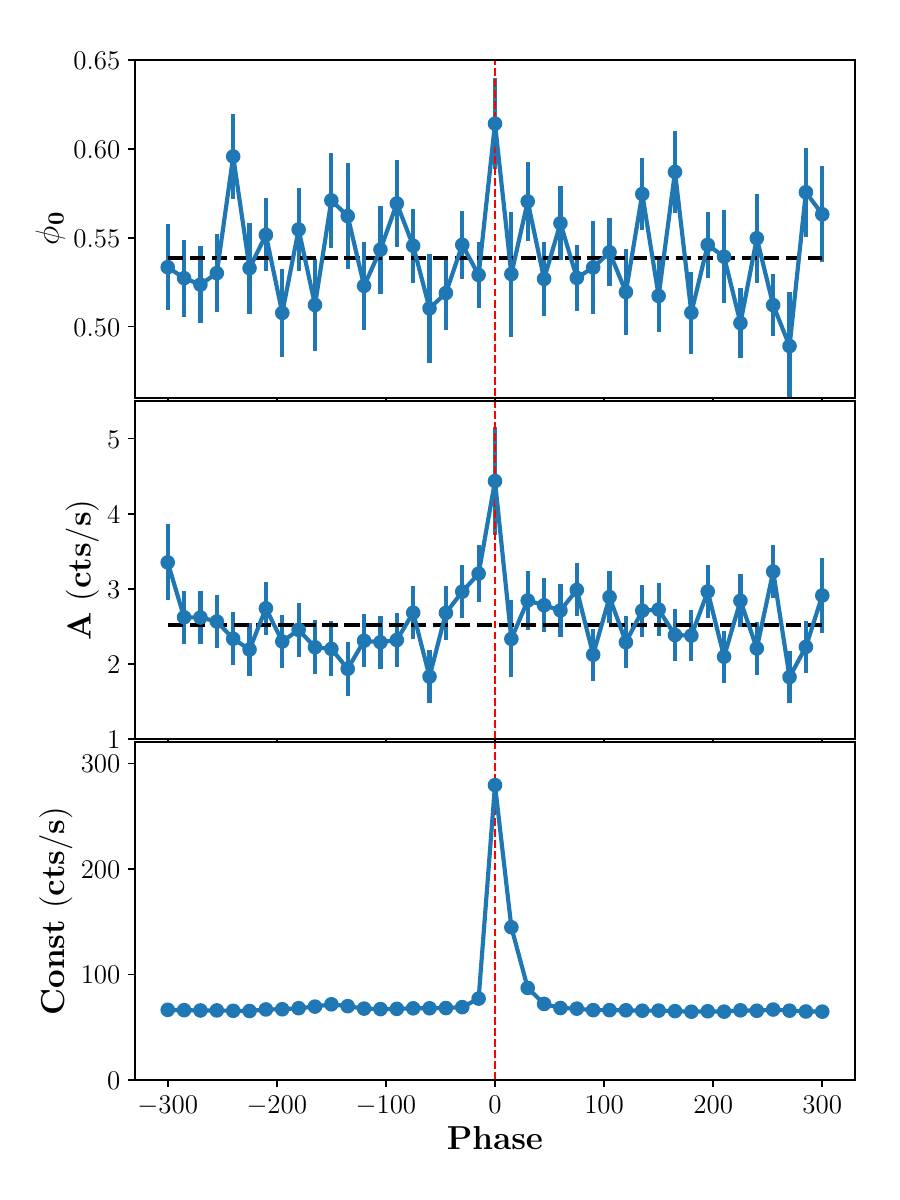}
\caption{
Parameters describing pulse profiles using a sinusoidal model. The red dashed line represents the time zero corresponding to the averaged burst peak.
}
\label{phase_A_const_ME}
\end{figure}

According to the burst and non-burst time intervals mentioned above, we folded events to obtain pulse profiles.
Here we adopted the ephemeris covering the whole outburst reported by \citet{LiZS2023}, where the timing solution can be described as a quadratic spin-up model.
Due to the statistical limitation, the pulse profile for each burst can not be used for comparison, and instead we only compare the averaged pulse profiles (shown in Figure~\ref{profile_compare}).
In general, there is no significant difference between the burst and non-burst pulse profiles except for the average count rate.
To describe the pulse profiles quantitatively, we fitted them with a sinusoidal function $f=A \sin(2\pi(\phi+\phi_0)) + const$, where $\phi$ is the spin phase, $A$ presents the pulsed modulation\footnote{$A$ is the absolute pulsed amplitude instead of the fractional amplitude.} and $const$ is the contribution from the unpulsed flux.
The results are listed in Table~\ref{tab1}, which indicates that only the $A$ parameter of ME increased by $(34\pm11$)\% during bursts.
We did not discover evident phase lags between burst and non-burst pulse profiles.
{On average, the phase lag is $(4\pm2)$\% cycles for HE and $(3\pm1)$\% cycles for ME.
But we caution that the averaging process may wash out small phase lags (see Discussion).
}
To investigate the evolution in details, we divided the ME data into many segments of 15 seconds and performed a time-resolved analysis (Figure~\ref{phase_A_const_ME}).
At the point corresponding the burst peak, $A$ is increased at a significance level of $\sim3\sigma$.
{This may indicate an enhancement of the pulsating persistent emission.
Alternatively, the increased $A$ could stem from nuclear-powered burst oscillations if they exist and are phase-locked to the persistent emission.
However, no burst oscillations were detected in the 15 bursts observed with \textit{NICER} \citep{Bult2022}.
}
We also studied this evolution with HE data, but we could not obtain decisive results due to large statistical errors.

\begin{table}
\centering
\caption{Fitting parameters of averaged pulse profiles of burst and non-burst intervals using a sinusoidal function.
{All uncertainties correspond to a confidence level of 68\%.}
}
\begin{tabular}{c|c|c|c}
\hline
                    &                  & burst                  & non-burst \\ \hline
\multirow{3}{*}{ME\,(8-30\,keV)} & $\phi_0$         &  0.57$\pm$0.01         &  0.54$\pm$0.01           \\ \cline{2-4} 
                    & $A$ (cts/s)      &  3.13$\pm$0.26         &  2.33$\pm$0.08           \\ \cline{2-4} 
                    & $Const$ (cts/s)  &  147.57$\pm$0.18       &  66.92$\pm$0.06          \\ \hline
\multirow{3}{*}{HE\,(30-100\,keV)} & $\phi_0$         &  0.59$\pm$0.02         &  0.55$\pm$0.01           \\ \cline{2-4} 
                    & $A$ (cts/s)      &  2.89$\pm$0.39         &  2.49$\pm$0.18           \\ \cline{2-4} 
                    & $Const$ (cts/s)  &  320.22$\pm$0.28       &  323.35$\pm$0.13         \\ \hline
\end{tabular}
\label{tab1}
\end{table}

\section{discussion}
In this paper, for the first time we were able to follow with {\it Insight}-HXMT the pulsations in the high energy band (30--100 KeV) where they should be caused by the persistent (i.e., accretion-powered) emission even during thermonuclear bursts.
Because the temperature of bursts is several keV, having negligible contributions at high energies, especially for above 30\,keV observed with \textit{Insight-HXMT}/HE detectors.
On the other hand, this means that hard X-rays are best suited to study the pulsed flux of the persistent emission during bursts in order to avoid the influence of burst oscillations.
During bursts, we found a deficit of hard X-rays, usually explained as the cooling of the corona by soft burst photons \citep[i.e.,][]{Chen2022}.
However, the periodic signal remains quite stable during these periods, different from the behavior of the non-pulsed emission.
This provides us with several direct conclusions as follows:
\begin{enumerate}[1)]
\item Hard X-rays in AMXPs consist of at least two independent components. One corresponds to the pulsed radiation emitted from the vicinity of the neutron star, which likely originates from the thermal Comptonization in the hot plasma near polar caps \citep{Gierlinski2005}.
The hot plasma is heated by the accretion shock by the infalling matter collimated by magnetic lines.
{Another component is the corona near the accretion disk, similar to other neutron star low mass X-ray binaries.
Since both components are produced by Comptonization processes, they are difficult to be distinguished based on spectral studies.
In our study, we find that the pulsed flux is almost unchanged, indicating that that the hot plasma near polar caps remains relatively stable.
This suggests that the hard X-ray deficit discovered by \citet{Chen2022} is primarily caused by by the cooling of the corona.
Therefore, if a burst is strong enough to completely cool down the corona, we could estimate the ratio of two Comptonization components by measuring the remaining hard X-ray during bursts.
}

\item 
The bursts observed in MAXI J1816-195 will not destroy or significantly change the accretion structure of polar caps.
This can be justified by the stable pulsed flux and phase during bursts (but see discussions below).
Up to now, it is still poorly known about the ignition location and how the flame propagates across the surface of the neutron star. 
%Convex studies of burst rising lightcurves
{Studying the convexity of the rising part of lightcurves} in MAXI J1816-195 suggests that the ignition may start near the equator (\citet{Chen2022,Maurer2008}, but see \citet{Goodwin2021}).
{If the magnetic confinement is not strong and the flame can cover polar caps}, bursts will provide as additional thermal seed photons being Comptonized by hot plasma near the neutron star surface.
This may be related to the increasing of the pulsed flux observed with \textit{Insight-HXMT}/ME.
However, detailed theoretical calculations are beyond the scope of this paper.
We note that bursts in MAXI J1816-195 are relatively faint, and their peak luminosities are only approximately $30\%$ Eddington luminosity since the ignition is in a hydrogen-rich environment \citep{Bult2022}.
On the other band, as for brighter bursts, especially those reaching the Eddington luminosity and the photospheric radius expansion (PRE) \citep{Galloway2021}, the interaction between burst photons and the infalling matter can not be neglected.
{In this case, the radiation pressure may hinder or inhibit the accretion process, leading to a decrease in the pulsed flux.}

\item On average, thermonuclear bursts observed in our sample will not significantly shift the position of polar caps in MAXI J1816-195.
This can be proven by the fact that no evident phase lag was observed during {thermonuclear} bursts.
Although it is believed that polar caps are mainly determined by the dipole magnetic poles of the neutron star, other factors can also affect their positions.
For example, in bursting pulsar GRO J1744-28, pulsations during type-II bursts (which are caused by spasmodic accretion instead of thermonuclear burning) lag behind their expected arrival times \citep{Stark1996,Koshut1998,Woods2000}.
They were explained as the accretion footprint pivoting on the neutron star surface when the accretion increased \citep{Miller1996}.
In MAXI J1816-195, the persistent emission only increases by 20\% during bursts due to the Poynting-Robertson drag \citep{Bult2022, Worpel2013}.
Therefore, a significant phase lag during bursts is not expected.
{In addition, our results are reminiscent of some nuclear-powered burst oscillations that are phase-locked to the accretion-powered pulsations \citep{Strohmayer2003, Watts2005,Watts2006,Watts2008,Cavecchi2022}.
As suggested by \citet{Cavecchi2022}, this rigid phase-locking is an averaging result, and for each individual burst, there is a repeating moderate ($\lesssim$0.1 cycles) phase drift.
We speculate that this averaging effect may also exist for accretion-powered pulsations during bursts, although burst oscillations and accretion-powered pulsations have different underlying mechanisms.
Therefore, a time-resolved study in hard X-rays with higher statistics is needed in the future.
Moreover, in XTE J1814-338 \citet{Strohmayer2003} discovered evident frequency drifts of burst oscillations only in bright bursts, suggesting that the phase-locking is related to the burst luminosity. 
Similarly, accretion-powered pulsations may also exhibit this behavior, and the phase lag should be investigated in brighter bursts in the future.
%On average, 
%burst oscillations of this source are coherent with the persistent pulsations
}
%The pulsations we reported here are powered by accretion, different from burst oscillations seen in soft X-rays. Given their different mechanisms, we do not know whether the frequency drift that occurs in phase-locked burst oscillations would be present in accretion-powered pulsations.

%which were reported in Type-II bursts of GRO J1744-28 \citep{Stark1996,Woods2000}.

%No phase shift during bursts was found in MAXI J1816-195.

\begin{acknowledgments}
This work is supported by the National Natural Science Foundation of China under grants No. 12173103, U2038101 and 12273030. 
This work is based on observations with \textit{Insight-HXMT}, a project funded by the China National Space Administration (CNSA) and the Chinese Academy of Sciences (CAS). This work is also supported by International Partnership Program of Chinese Academy of Sciences (Grant No.113111KYSB20190020).
\end{acknowledgments}

\appendix 
\renewcommand\thefigure{\thesection.\arabic{figure}}    
\setcounter{figure}{0}  
\section{The influence of the HE background}
{The instrumental background of {\it Insight}-HXMT/HE is mainly related to its attitude and the orbit. 
We performed many simulations by assuming random reference times to extract lightcurves from {raw events of actual observations}, as we did with real bursts' time. 
We found that despite averaging 73 lightcurves, the long-term variability of the background could not be fully mitigated.
In Figure~\ref{simulation}, we show a representative example of averaged lightcurves which exhibits a downward trend.}

{
In addition, we also tested the downward trend by analyzing a sample of bursts observed during low-background orbits.
This sample comprises 12 bursts, for which we were able to estimate the background using the official tool \textit{hebkgmap} and generate an averaged lightcurve with background subtraction (shown in Figure~\ref{good_sample}).
Clearly, the hard X-ray deficit is still significant around the burst peak, while the overall downward trend disappears.
This concludes that the downward trend shown in Figure~\ref{ME_HE_lc} is likely due to an average change of the instrumental background.
}

\begin{figure}
\centering
\includegraphics[width=0.6\textwidth]{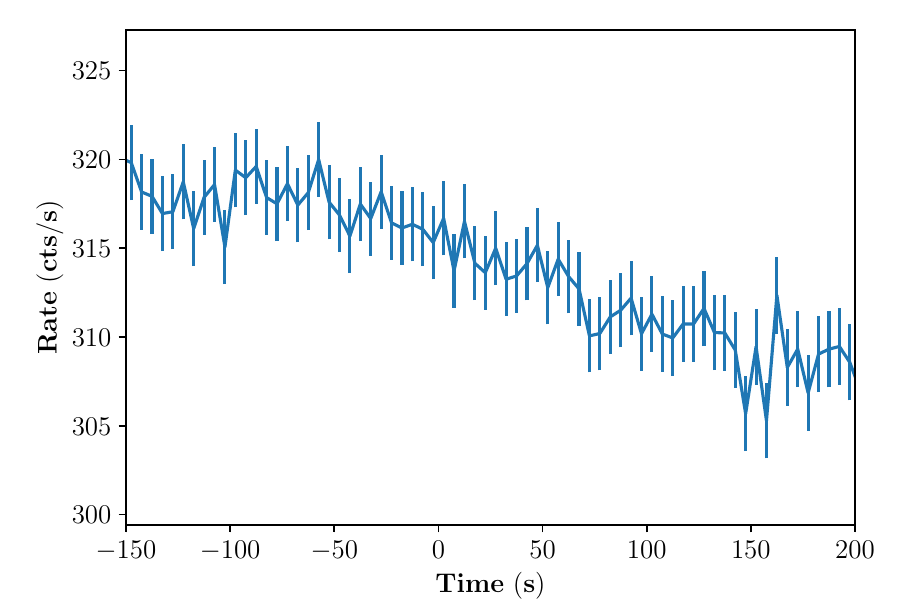}
\caption{A representative example of averaged simulated lightcurves.
This suggests the long-term variability of the HE background can not be fully eliminated by stacking lightcurves.
}
\label{simulation}
\end{figure}

\begin{figure}
\centering
\includegraphics[width=0.6\textwidth]{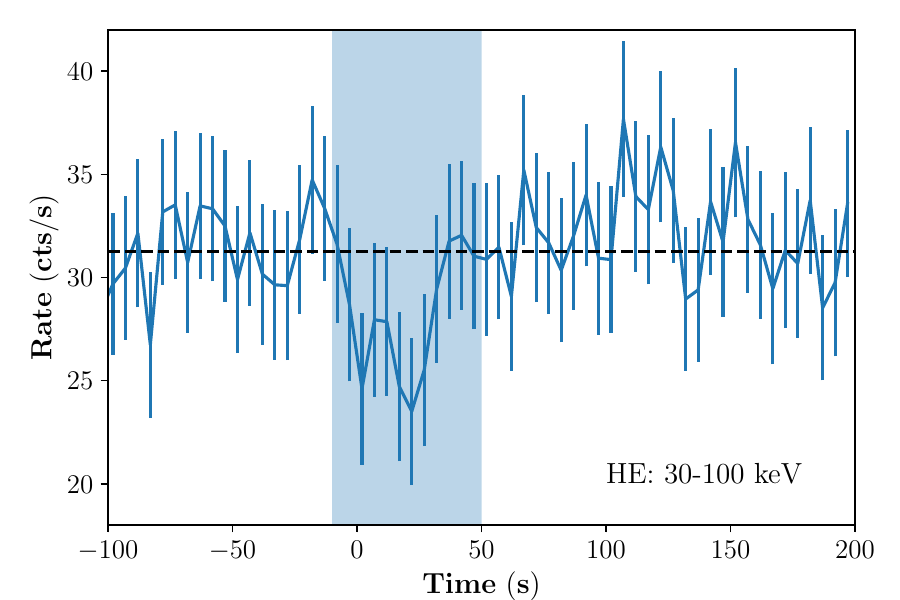 }
\caption{
Averaged background-subtracted lightcurves of 12 thermonuclear bursts observed in the low-background regions of {\it Insight}-HXMT/HE. The shaded region represents the burst interval.
}
\label{good_sample}
\end{figure}

\end{enumerate}

%It was difficult to be studied because of the small number of bursts in AMXPs. and the strong contamination of burst photons.

\bibliography{sample631}{}
\bibliographystyle{aasjournal}

%% This command is needed to show the entire author+affiliation list when
%% the collaboration and author truncation commands are used.  It has to
%% go at the end of the manuscript.
%\allauthors

%% Include this line if you are using the \added, \replaced, \deleted
%% commands to see a summary list of all changes at the end of the article.
%\listofchanges

\end{document}